\documentclass[12pt, a4paper]{article}
\usepackage{latexsym}

\title{\large\bf The basic principles of geometrization\\
 of the quantum mechanics}
\author{{\large K.B. Korotchenko}\\
{\small\bf Tomsk Polytechnical University, Rossia}\\
{\small\bf e-mail:  kost@phys.dfe.tpu.edu.ru}}
\date{}
\begin{document}
\maketitle
\begin{abstract}
The basic principles of the quantum mechanics in the
$K$-field formalism \cite{p1} are stated in the paper.
$K$-field formalism arises from geometric generalization of
de Broglie postulate. So, the quantum theory equations
(including well-known Schr\"{o}dinger, Klein-Gordon and
quadratic Dirac equations) are obtained as the free wave
equations on a manifold metrizing force interactions of
particles.

In this paper, describing wave properties of particles we
will restricted ourself with construction special geometric
formulation of force interactions.
\end{abstract}

n1. The de Broglie hypothesis initial sense is that it is
possible to spread the formulas describing a photon behaviour
to mass particles too. Such a view on the de Broglie theory
allows one to use it for a formal description of a
microparticles behaviour in terms of a photon behaviour
formal description.

Such a conception of the de Broglie hypothesis we shall
accept as a basis of the quantum theory formalism.

n2. To present correlation of photon and microparticle
description more obviously let's make a table (see tab. 1).

Obviously by virtue of such essential difference in
description of photons and microparticles the direct
prolongation of the equations for photons on mass particles
is impossible.

Let's look at this problem a little differently. Let's
consider motion of particles from an isolated observer's view
point.

Photons move along an isotropic geodesic lines of the
Minkowski four-space $V_4 $. Hence, it is necessary to build
a four-space in which mass particles should move (relatively
to an isolated observer) along an isotropic geodesic lines of
this space.
\begin{center}
\begin{tabular}{||p{62 true mm}|p{62 true mm}||}
\multicolumn{2}{r}{Table 1}\\ \hline
\multicolumn{1}{||c|}{photons} &
\multicolumn{1}{c||}{mass particles} \\
\hline \hline
move uniform rectilinearly with the velocity $c $ relatively
to any inertial frame & move along any trajectories with any
velocities (less than $c $) relatively to any inertial
frame\\ \hline the wave properties are described by the
equation of d'Alember
\[ \Box A^{\mu}\: =\: 0\: \]
for an electromagnetic field four-potential $A^\mu $ & the
wave properties are described by the Klein-Gordon equation
\[ [(E - V)^2 + \hbar^2 c^2 \triangle - m^2 c^2] \Psi\: =\: 0 \]
for a particle $\Psi$-function \\ \hline \hline
\end{tabular}
\end{center}
\ \\
Let's now collect all the isotropic surfaces of separate
observers in one four-manifold $^kV_4 $. We shall obtain the
manifold $^kV_4 $ in which mass particles will move
(relatively to any observer) along isotropic geodesic lines
of this manifold for any motion.

Hence, description of a mass particle on the manifold $^kV_4
$ and description of a photon in the Minkowski four-space
$V_4 $ becomes equivalent. And so, we can formulate the de
Broglie hypothesis as follows:
\begin{itemize}
\item[\ ] to describe wave properties of particles it is necessary
\begin{itemize}
\item[\ -] to build the manifold $^kV_4 $ in which mass particles
 move along geodesic lines of this manifold for any motions;
\item[\ -] to build an operator the similar to that d'Alember on the
 manifold $^kV_4 $
 (so-called the de Rham operator $^{(k)}\triangle $);
\end{itemize}
\item[\ ]
 then the equation
\begin{equation}
\label{e01}
 ( ^{(k)}\triangle k )_{\mu}\: =\: 0\: ,
\end{equation}
 where $k_{\mu}$ is the $K$-field potential should describe
 the wave properties of mass particles.
\end{itemize}
 Let's sum up everything, mentioned above, in a table (see tab. 2)
\begin{center}
\begin{tabular}{||p{62 true mm}|p{62 true mm}||}
\multicolumn{2}{r}{Table 2}\\
\hline
\multicolumn{1}{||c|}{photons} &
\multicolumn{1}{c||}{mass particles} \\
\hline
\hline
move along isotropic geodesic lines of the Minkowski
four-space $V_4 $ & move along isotropic geodesic lines
 of the manifold $^kV_4 $\\
\hline
 the wave properties are described by the equation of d'Alember
\[ \Box A^{\mu}\: =\: 0\: \]
 for the electromagnetic field four-potential $A^\mu $ &
 the wave properties are described by the equation
\[( ^{(k)}\triangle k )_{\mu}\: =\: 0\: \]
 for the $K$-field potential $k_\mu $, where $^{(k)}\triangle $
 is the de Rahm operator and $k_{\mu} $ is linear form \\
\hline
\hline
\end{tabular}
\end{center}

n3. In this paper, describing wave properties of particles we
will restricted ourself with construction special geometric
formulation of force interactions.

Geometrization of an interaction consists in finding a metric
space in which the test particle trajectories are geodesic
lines \cite{p2}. This is the starting point of Einstein
concept of geometrization.

An interesting method of metrization of arbitrary force
interactions corresponding to this concept was presented in
\cite{p3}. In this method of metrization, the test particles
move along geodesic lines.  However, the force fields are
related with the components of the connection tortion tensor
of a pseudo-Euclidean space.  In this sense, the metrization
of force interactions presented in \cite{p3} does not
correspond to the Einstein concept because the metric
properties of the space do not depend on force fields.

So, we shall consider a metric statement of force
interactions in which, as in \cite{p3}, the test particle
motion equations represent a special form of Newton's second
law in four-dimensional form but the metric tensor and
physical fields are interdependent.

To avoid the problems connected with the distinction between
the concepts of a reference frame and a coordinate system
\cite{p3}, different observers (i.e., reference frames) will
be associated with different isotropic surfaces on the
manifold $^kV_4 $.

n4. The states of the test particles (of mass $m $ and charge
$e $) in potential fields will be called classical states.
Correspondingly, all the characteristics of the particle
describing its behavior in the classical state (trajectory,
velocity, momentum, energy, etc.) will be called classical.

It should be emphasized that all classical characteristics
should be measured relatively to one specific reference
frame. Any inertial frame (IF) may be selected as that
reference frame.

Let's consider some a four-dimensional space with the metric
\begin{equation}
\label{e1}
 ^{(k)}dS^2\: =\: {^{(k)}}g_{oo}(x^i, t) c^2 dt^2\: +
               \: g_{ik} dx^i dx^k\: ,
\end{equation}
where ($- g_{ik} $) is the metric tensor of the Euclidean space
$V_3 $.

Any classical trajectory $x^i = x^i (t) $ may be considered
as a line defined by the equation ${^{(k)}}g_{oo}(x^i (t),
t)c^2 dt^2 =
 -- g_{ik} dx^i dx^k $. And so along the line
\begin{equation}
\label{e2}
 {^{(k)}}g_{oo}(x^i (t), t)\: =\: v^i v_i/c^2\: ,
\end{equation}
where $v^i $ is the particle velocity measured relatively to the
specified IF ($v_i = -- g_{ik} v^k $).

Thus, each point $p $ of the classical particle trajectory
$x^i = x^i (t) $ in $V_3 $ may be considered as a line lying
on the isotropic surface $^kG_{o3} \subset {^k}V_4 $
described by the equation ${^{(k)}}g_{oo}c^2 dt^2 = -- g_{ik}
dx^i dx^k $.

Hence, each point $p \in V_3 $ may also be considered as a
point of the isotropic surface $^kG_{o3} $ in $^kV_4 $. That
is an isotropic surface $^kG_{o3} \subset {^k}V_4 $ may be
constructed at points of space $V_3 $. By changing the values
of the initial parameters, a set of points covering the whole
of $^kG_{o3} $ may be obtained. And by transiting from one
reference frame to another, a set of surfaces $^kG_{o3} $
covering the whole of $^kV_4 $ may be obtained. That is an
imbedding \cite{p2} (enclosure in a space of higher
dimensionality) may be constructed.

n5. According to Eq.(\ref{e1}), the method of enclosure
described in Sec.5 should have the distinctive property.
Namely, the geometry of the enclosing space $^kV_4 $ should
have no influence on the geometric properties of the enclosed
space $V_3 $ (should not change the metric tensor $g_{ik} $).
In other words, the imbedding must occur at those points of
$^kV_4 $ at which the external curvature of the enclosed
surface is zero.

So, then it follows from the Gauss--Vaingarten equations (see
\cite{p4}, for example), the absolute differential of the
space $^kV_4 $ (denoted by ${^{(k)}}\nabla (\cdots) $) is
defined by the equation
\begin{equation}
\label{e3}
 ^{(k)}\nabla A^{\mu}\: =\: ({^{(3)}}\nabla_i A^{\mu}) dx^i\: +
                       \: ({^{(4)}}\nabla_o A^{\mu}) dx^o\: .
\end{equation}
     Equation (\ref{e3}) may also be rewritten in the form
\begin{equation}
\label{e4}
 ^{(k)}\nabla A^{\mu}\: =\: {^{(k)}}D A^{\mu}\: +
     \: {^{(k)}}\Gamma^{\mu}_{\nu o} A^{\nu} dx^o\: ,
\end{equation}
where ${^{(k)}}DA^i = DA^i + {^{(k)}}S^i_{k l}A^k dx^l $
is the absolute differential of the Euclidean space
$V_3 $ (here ${^{(k)}}S^i_{k l} = S_{k l}{^i}
- S_l{^i}{_k} - S_k{^i}{_l} $
and $S_{k l}{^i} $ is the tortion tensor) and
${^{(k)}}\Gamma^{\mu}_{\nu o} $ is the connection of the space
$^kV_4 $.

n6. To obtain a more detailed description, the definition of
the absolute differential $^{(k)}\nabla (\cdots) $ is written
in standard form \cite{p2}, \cite{p5}
\begin{equation}
\label{e5}
 ^{(k)}\nabla A^{\mu}\: =\: (\partial_{\nu} A^{\mu}\: +
     \: \Gamma^{\mu}_{\omega \nu} A^{\omega}) dx^{\nu}\: ,
\end{equation}
where $2\Gamma^{\mu}_{\omega \nu} =
2 {^{(k)}}\Gamma^{\mu}_{\omega \nu} +
Q^{\mu}_{\omega \nu} $,
at that $Q^{\mu}_{\omega \nu} =
{^{(k)}}g^{\mu \gamma} ({^{(k)}}Q_{\omega\nu\gamma} +
{^{(k)}}Q_{\nu\gamma\omega} -
{^{(k)}}Q_{\gamma\omega\nu}) $ and
${^{(k)}}Q_{\mu\nu\omega} =
- {^{(k)}}\nabla_{\mu} \Big({^{(k)}}g_{\nu \omega} \Big) $;
${^{(k)}}\Gamma^{\mu}_{\omega \nu} =
\Big\{^\mu_{\omega\nu} \Big\} +
{^{(k)}}S^{\mu}_{\omega\nu} $,
where $\Big\{^\mu_{\omega\nu} \Big\} $ is the Christoffel symbol
and ${^{(k)}}S^{\mu}_{\omega\nu} = S_{\omega\nu}{^{\mu}}
- S_{\nu}{^{\mu}}{_{\omega}} - S_{\omega}{^{\mu}}{_{\nu}} $
at that $S_{\omega\nu}{^{\mu}} $ is the tortion tensor.

If it is required that the definition in Eq.(\ref{e5})
coincide with that in Eq.(\ref{e4}), the result obtained is
\begin{equation}
\label{e6}
 2{^{(k)}}\Gamma^i_{o j} dx^j\: +\:
    Q^i_{o \omega}dx^{\omega}\: =\:
    Q^i_{j \omega}dx^{\omega}\: =\:
 2{^{(k)}}\Gamma^o_{\mu j} dx^j\: +\:
    Q^o_{\mu \omega} dx^{\omega}\: =\: 0\: ,
\end{equation}
which must be satisfied if the imbedding described in Secs.5
and 6 is possible.

It may readily be demonstrated that the absolute differential
$^{(k)}\nabla (\cdots) $ of space $^kV_4 $ defined by
Eq.(\ref{e4}) describes a nonmetric transfer in $^kV_4 $. In
fact
\begin{equation}
\label{e7}
    {^{(k)}}Q_{o o o}\: =\:
   2{^{(k)}}g_{o o}{^{(k)}}S^o_{o o}\: ,\; \;
    {^{(k)}}Q_{i o o}\: =\: - \partial_i {^{(k)}}g_{o o}\: .
\end{equation}
The remaining ${^{(k)}}Q_{\mu\nu\omega} = 0 $. As a result
the equations (\ref{e6}) take the form
\begin{eqnarray}
\label{e8}
 {^{(k)}}S^o_{o j} dx^j & = & - \Big\{^o_{o j}\Big\} dx^j -
 2{^{(k)}}S^o_{o o} dx^o\: ,\nonumber \\
\label{e8-1}
 {^{(k)}}S^o_{i j} dx^j & = & -  \Big\{^o_{i j}\Big\} dx^j +
 \Big\{^o_{i o}\Big\} dx^o\: ,\\
\label{e8-2}
 {^{(k)}}S^i_{o j} dx^j & = & -  \Big\{^i_{o j}\Big\} dx^j +
 \Big\{^i_{o o}\Big\} dx^o\: ,\nonumber
\end{eqnarray}
Hence it is clear that the tortion $S_{\omega\nu}{^{\mu}} $
is nonzero.

Thus, the imbedding described in Sec.2 generates in $^kV_4 $
a geometry with tortion and a nonzero covariant derivative of
the metric tensor.

n7. The test particle motion equations are now considered. It
is desirable for these equations to coincide with the
geodesic equations in $^kV_4 $. Then these equations should
take the form
\begin{equation}
\label{e9}
D p^{\mu}\: =\: - ^{(k)}\Gamma^{\mu}_{\nu o} p^o dx^{\nu}\: ,
 \; \; (p^{\mu}\: =\: m dx^{\mu}/d\tau)\: .
\end{equation}
Taking this into account, the condition
$dx_i dp^i = dx^o dp^o $ leads to the equation
\begin{equation}
\label{e10}
 {^{(k)}}S^j_{\nu o} dx^{\nu} dx_j\: =\:
 \Big(\Big\{^o_{\nu o}\Big\} + {^{(k)}}S^o_{\nu o}\Big)
 dx^{\nu} dx^o - \Big\{^j_{\nu o}\Big\} dx^{\nu} dx_j\: ,
\end{equation}
which, together with Eq.(\ref{e8}), describes all the
nonzero components of ${^{(k)}}S^{\mu}_{\omega\nu} $.

Further, it is readily evident that, if the components
${^{(k)}}S^o_{\nu o} $ are chosen in the form
\begin{equation}
\label{e11}
 {^{(k)}}S^o_{\nu o}\: =\:
 [\partial_{\nu} \ln ((1 - {^{(k)}}g_{o o})/{^{(k)}}g_{o o})]/2\: ,
\end{equation}
the four-momentum $p^o $ component is found to be
\begin{equation}
\label{e12}
 p^o\: =\: C_1(1 - {^{(k)}}g_{o o})^{-1/2}\: ,
\end{equation}
where $C_1 = const $. Assuming that $C_1 = m c $, it is found
that $d\tau = (1 - {^{(k)}}g_{o o})^{-1/2} dt $.

Hence, Eqs.(\ref{e10}) and (\ref{e11}) are the necessary and
sufficient conditions for the motion equations (\ref{e9}) to
be noncontradictory.

Thus, the classical particle trajectories in the potential
fields specified with respect to a definite IF may be
represented as geodesic lines lying on isotropic surfaces of
some configurational space $^kV_4 $ the connection of which
has tortion, while the transference is nonmetric. The
geometry of the space $^kV_4 $ has the distinctive property
that the magnitude of the nonmetricity of the transfer and
the tortion are determined by specifying the metric
coefficient ${^{(k)}}g_{o o} $ under the condition that the
mixed components ${^{(k)}}g_{o i} \equiv 0 $.

\end{document}